\documentstyle[aps,preprint,epsfig]{revtex}

\begin{document}
\title{Size Matters: Origin of Binomial Scaling in Nuclear Fragmentation
       Experiments}
\author{Wolfgang Bauer and Scott Pratt}
\address{Department of Physics and Astronomy and \\
National Superconducting Cyclotron Laboratory,
\\Michigan State University, East Lansing, MI 48824~~USA}
\date{\today}
\maketitle

{\abstract The relationship between measured transverse energy, total
charge recovered in the detector, and size of the emitting system
is investigated.  Using only very simple assumptions, we are able to
reproduce the observed binomial emission probabilities and their
dependences on the transverse energy.}


During the last decade, evidence has been mounting that nuclear matter
undergoes a phase transition in the nuclear fragmentation process.
From general consideration regarding the elementary nucleon-nucleon 
interaction (repulsive at short and attractive at intermediate
distances), we expect the nuclear phase-diagram to show a Van-der-Waals
``liquid-gas'' phase transition of first order, terminating in a
second-order transition at the critical point.

Recent observations point to evidence of first and second order
transitions.  In experiments studying Au-Au collisions 
conducted at the GSI, a measurement of
the temperature as a function of excitation energy found possible evidence
for a two-phase coexistence regime \cite{Poc95}, not unlike the
scenarios predicted by statistical multifragmentation models with
excluded volume \cite{Gro90,Bon95}.  Other experiments conducted at
the Bevalac focussed on the extraction of critical exponents from
(almost) completely reconstructed Au-fragmentation events on C-targets,
studying the dependence of the second moment of the charge distribution and
size of the largest fragment as a function of the total charged particle
multiplicity \cite{Gil94,Ell94,Rit95}.  It was 
shown \cite{BF95,BB95} that these data
are consistent with the second-order phase transition predicted
by the nuclear percolation model \cite{Bau85,Bau86,Bau88,Cam86}.

If one wants to gain a fundamental understanding of the fragmentation
process that goes beyond simple equilibrium
model descriptions of the phenomena, then a proper description of the
origin and time evolution of fluctuations is 
essential \cite{RK81,BBD87,Cho96,Gha98}, in particular if one wants
to understand why particular molecular dynamics codes produce
fragments (or not!), and what their connections to the fundamental
processes of nuclear fragmentation 
are \cite{AS86,BG88,Fel90,Aic91,Ono92,Ono92b,Lat94,Pra95,FS97,OR97,KD97}.

In this light, the recent findings of Moretto {\it et al.} are all the more
surprising \cite{Mor93b,Mor95,Pha96,Mor97,Bea98}.  This group found
that the probability $P_n$ of emitting $n$ intermediate
mass fragments (IMFs) follows a binomial distribution
\begin{equation}
      P_n(m,p) = \frac{m!}{n!\,(m-n)!}p^n(1-p)^{m-n}
\end{equation}
The parameters $m$ and $p$ are related to the average and variance of
the distribution.
\begin{eqnarray}
      \langle n\rangle &=& \sum_{n=0}^{\infty}n\,P_n(m,p) = m\cdot p\\
      \sigma^2 &=& \sum_{n=0}^{\infty}(n-\langle n\rangle)^2\,P_n(m,p) =
          m\,p\,(1-p)
\end{eqnarray}
This result suggest that one may interpret the parameter $p$ as the elementary
probability for the emission of one fragment and the parameter $m$ as the total
number of tries.  This would indicate that the problem of multi-fragment
emission is reducible to that of multiple one-fragment emission.
The claim for reducibility and its interpretation as the consequence
of a simple barrier penetration phenomenon 
was further strengthened by the observation
that $\ln(p^{-1})$ has a linear dependence on $1/\sqrt{E_t}$, where $E_t$ is
the total transverse energy, $E_t = \sum_l E_{kl}\sin^2\theta_l$.  
Finally, the same scaling was found for different
beam energies and different projectile-target combinations.

Other authors have criticized the above work, pointing our that there are
different emission probabilities for different size IMFs, that there are
problems in the transformation between the total transverse energy and a
true thermal energy \cite{Tok97}, and focussing on autocorrelations between the
number of IMFs and the transverse energy \cite{TD98}.  (See also the replies
to these criticisms in refs.\ \cite{Bea98,Mor98}.)

In the present note we add to this discussion by showing how binomial
distributions arise naturally from finite size effects.  In particular,
we focus on the dependence of the experimentally recovered charge as
a function of the measured transverse energy.   We then demonstrate
why the dependence of the binomial parameter $p$ on the total
transverse energy arises.

We begin our study by generating power-law distributed random
fragmentation events.  This is accomplished by determining the charge
of individual fragments with a probability distribution proportional
to $Z^{-\tau}$, where $Z$ is the fragment charge, and $\tau$ is the power-law
exponent.  For definiteness, we wish to generate events with exactly
$Z_{\rm sys}$ charges.  If an event has less than $Z_{\rm sys}$ 
charges, we add another fragment;
if it has more than $Z_{\rm sys}$ charges, we throw it out.  
For an infinite system, 
we would expect the multiplicity distributions for individual fragments
of a given $Z$ to follow a Poisson distribution,
\begin{equation}
      Q_n(\lambda) = \frac{\lambda^n\,\exp(-\lambda)}{n!}\ \ \
      {\rm with:}\ \ \ \lambda = \langle n\rangle = \sigma^2
\end{equation}
And since the combined
probability distribution of two Poisson-distributed variables is again
a Poissonian,
\begin{equation}
      Q_n(\lambda_1) \otimes Q_n(\lambda_2) \equiv
      \sum_{i=0}^n Q_i(\lambda_1)\cdot Q_{n-i}(\lambda_2)
      = Q_n(\lambda_1+\lambda_2)\ ,
\end{equation}
we would expect that the multiplicity distribution of the total number
of intermediate mass fragments (IMFs) is also Poissonian.

The individual probability distributions for IMFs,
however, cannot be exactly Poissonian, because the tails of the distributions
are cut off due to the finite size of the emitting system.  Thus the 
probability distributions in our simulation are closer to a binomial
distribution with rather large values of $m$ and small values of $p$.  
(When $m\rightarrow\infty,\ p\rightarrow 0$ such that $mp=$const., we obtain
a Poissonian as the limit of a binomial distribution.)
Typical values of $p$ we find for the probability distributions of our
individual fragments are $\le 3\cdot 10^{-2}$ for a system of 100 total
charges.  These small values of $p$ imply that the probability distributions
are very close to a Poisson distribution.

We now ask what the combined probability distribution for fragments charges
in the interval 3 to $k$, $k=4,5,...,30$ is.  
(If we use $k=20$, this corresponds to the usual definition of IMFs.)
We find numerically that
to very good approximation this distribution is again a binomial distribution,
for all values of $k$.  The binomial parameters $p_k$ and $m_k$ have a
monotonical behavior as a function of $k$:  $p_k$ rises monotonically until
it saturates at $k = Z_{\rm sys}$, and $m_k$ falls monotonically.  
This results directly from
the mathematical fact that the mean of the combined probability distribution
is the sum of the mean values of the individual distributions, but the
variance is always smaller than the sum of the variances for the folding
of binomial distributions.  

In figure 1, we show the behavior of the
parameters $p_k$ and $m_k$ of the combined probability distributions 
as a function
of $k$, the upper limit charge for the folding procedure, for different values
of $Z_{\rm sys}$.  For each value of $Z_{\rm sys}$, we generated $10^4$ events.
This figure already contains the essential key to 
understanding the patterns observed by Moretto and collaborators.
We can clearly see that as we include more and more fragments in the definition
of IMFs the extracted values of the binomial parameter $p_k$ increase, and
those for $m_k$ decrease.  
We can also see that $p_3$ ($= p_k(k$=3$)$)
decreases as we
increase the total charge of the fragmenting system.  This is expected:
the larger the total available charge, the closer the probability distributions
of individual fragments (in this case $Z=3$) will be to the Poissonian limit.
It is essential to note that the values of $p_k$ for each system size
saturate at $k=Z_{\rm sys}$.  This is obvious, because we cannot have IMFs
larger than the total charge available.  But this obvious fact has an 
interesting consequence: The smaller the system size, the fewer the terms
that can contribute to the construction of the asymptotic value of $p_k$,
and the lower the asymptotic value of $p_k$.  

This fact, combined with
the dependence of $Z_{\rm sys}$ on the transverse energy, already is the
explanation for the scaling observed by Moretto {\it et al}.

What is the dependence of $Z_{\rm sys}$ on the transverse energy, $E_t$, in
the experiments of Moretto {\it et al.}?  This is shown in fig.\ 2 for the
reaction Kr+Au at 55 $A\,$MeV \cite{Tsang,Wil97}.  The filled plot symbols
show the mean $Z_{\rm sys}$ for each value of the $E_t$, and the error bars
give the width (standard deviation) of the distribution.  The dominant feature
of this figure is the linear rise of the mean value of  $Z_{\rm sys}$ with
$E_t$,
\begin{equation}
      \langle Z_{\rm sys}(E_t)\rangle \approx 2 + 0.092\ E_t/{\rm MeV}
\end{equation}
for values of $E_t$ less than 0.7 GeV, and the saturation of $Z_{\rm sys}$ for
larger values.

The width of the $Z_{\rm sys}(E_T)$-distribution is 
significant, on the order of
10 units of charge.  If we wish to construct the probability distributions
of intermediate mass fragments by using our knowledge of the dependence
of the binomial parameters $p$ and $m$ on the system size (fig.\ 1), and the
dependence of the system size on transverse energy, we have to integrate
over the experimentally measured 
width of the $Z_{\rm sys}(E_T)$-distribution.  The resulting values
of $p$ for the integrated distributions are shown as a function of $E_t$
in fig.\ 3.  We have run three different calculations, using three different
values of the exponent $\tau$
in our fragment production probability distributions,
$p(Z)\propto Z^{-\tau}$.  In this figure, we display the results in the same
way that Moretto {\it et al.} have done.  One can already see that there
is qualitative agreement with the tendencies observed by the Moretto-group:
for all values of $\tau$, we observe an approximately linear rise of 
$\ln p^{-1}$ with $1/\sqrt{E_t}$.  This, however, is {\em not} the consequence
of some kind of thermal scaling.  Instead, it is purely a consequence
of the variation of the size of the emitting system as a function of the
transverse energy, and with it a change in the effective parameter $p$
in the binomial probability distribution.

The projectile and target masses only enter into our consideration as upper
cutoffs for the possible maximum values for $Z_{\rm sys}$, and with it the
upper values of $E_t$.  The experimentally found functional dependence
of $Z_{\rm sys}$ on $E_t$ that we show in fig.\ 2 for the system Kr+Au 
is basically the same for all target-projectile combinations and beam
energies below approximately 100 $A\,$MeV (for higher beam energies, we have
sizeable radial flow contributions to the transverse energy).  
Transverse energy basically measures impact parameter, that is to say
size of the emitting system; and for any experiment measuring inclusive
fragment distributions the results will be very much similar to the
ones displayed in fig.\ 2.
This explains
the universal scaling observed by Moretto {\it et al.} without the need
for invoking some deeper reason for this apparent universality; a plot
of $\ln p^{-1}$ vs.\ $1/\sqrt{E_t}$ is dominated by effects of the
variation of the size of the emitting system.
We should point out that our findings are not dependent
on the fact that the intermediate mass fragments carry transverse energy
themselves.  The only correlations entering our analysis are the
experimental ones between transverse energy and system size.

We can obtain more-or-less
complete agreement with the experimental data, if we allow
the power-law parameter $\tau$ for the fragment mass distribution to vary with
impact parameter and with beam energy.  This variation
is a well-documented experimental fact \cite{Wil97,Li93,Li94}; 
the experimentally
observed value of $\tau$ increases with impact parameter and therefore
falls with transverse energy.  If we assume
\begin{equation}
      \tau(E_t) = 3.5 - E_t /(0.5\ {\rm GeV})\ ,
\end{equation}
then we get the result displayed in fig.\ 4.  Our calculations are represented
by the plot symbols.  The error bars are statistical and computed on the basis
of $2\times 10^4$ events for each point.  The solid line is a fit to
the experimental results of Moretto {\it et al}.  As one can see, there is
very good agreement.  Assuming other functional dependences of $\tau$
of $E_t$ may even yield better results.
This agreement, however, is not quite as relevant as
the main message we wish to impress on the reader:  The universal scaling
of $\ln p^{-1}$ vs.\ $1/\sqrt{E_t}$ is almost exclusively due to the finite
size of the system emitting the fragments and the dependence of 
the measured value of the transverse energy on that size.

Even though the main message of the present note is that the $E_t$-dependence
of the extracted binomial parameters of 
the fragment multiplicity distributions can be explained rather 
straightforwardly, we do not wish to convey the message that there
is no interesting information that one can extract from this type of
analysis.  For instance, the effects of varying system size could be
eliminated with utilization of completely reconstructed fragmentation
events.  For these types of events, percolation models predict
a transition between sub- and super-Poissonian fluctuations near the
percolation threshold \cite{Gha98}.  
This type of behavior is not expected in a
sequential model.   Once the kind of correlations discussed by us
above are removed, then this type of fluctuations analysis should
yield insightful information about the character of the nuclear fragmentation
phase transition.

\acknowledgments{This work was supported by the National Science Foundation,
grant PHY-9605207.}

\clearpage

\begin{figure}
\vspace*{15cm}

\includegraphics{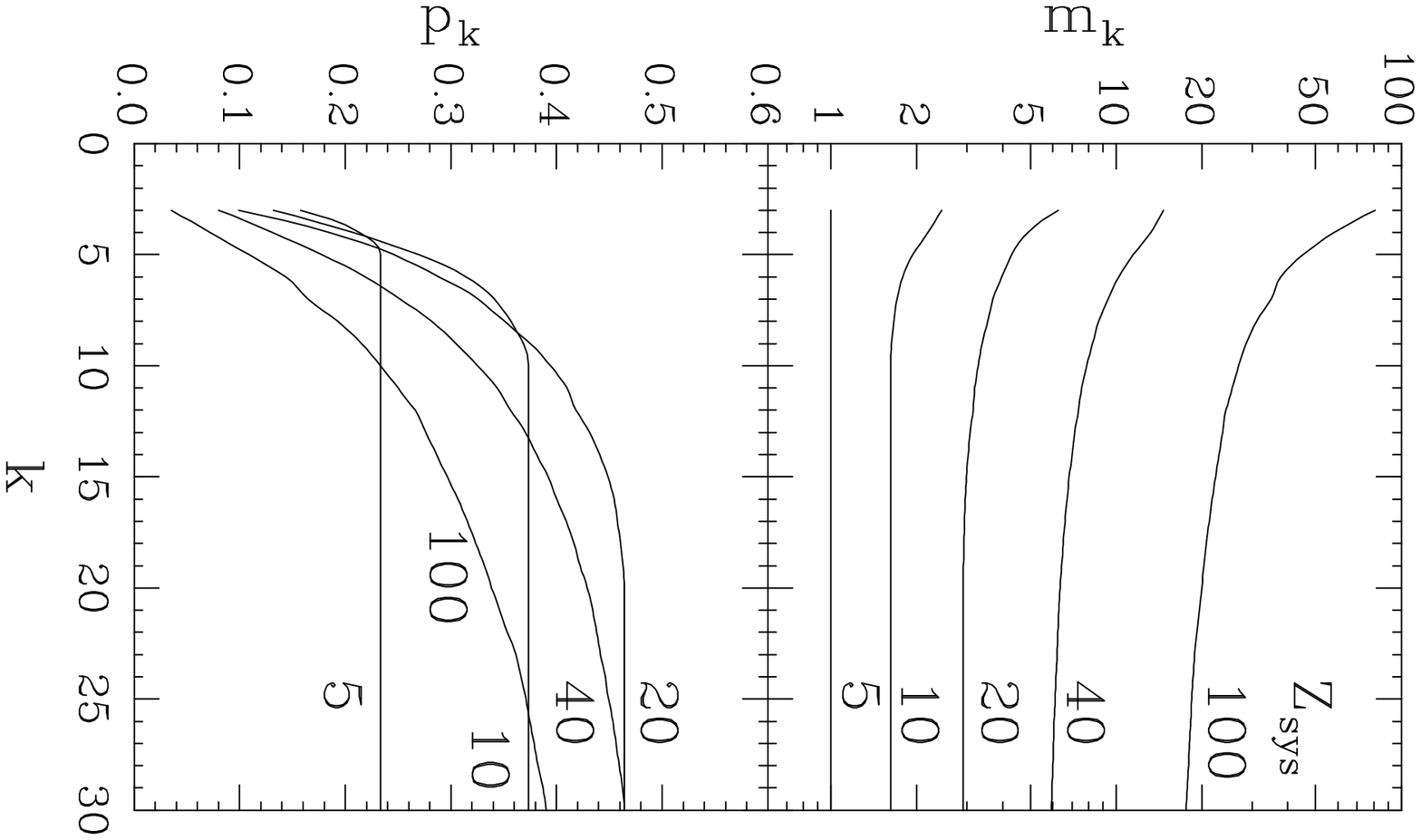}

\vspace*{3cm}

\caption{Dependence of the binomial fit parameters $p_k$ and $m_k$ of
         the probability distributions for intermediate mass fragments
         from charge $Z=3$ to $Z=k$ on the upper summation
         limit $k$, for different total charges of the fragmenting system, 
         $Z_{\rm sys}$.}
\end{figure}

\clearpage

\begin{figure}
\vspace*{15cm}

\includegraphics{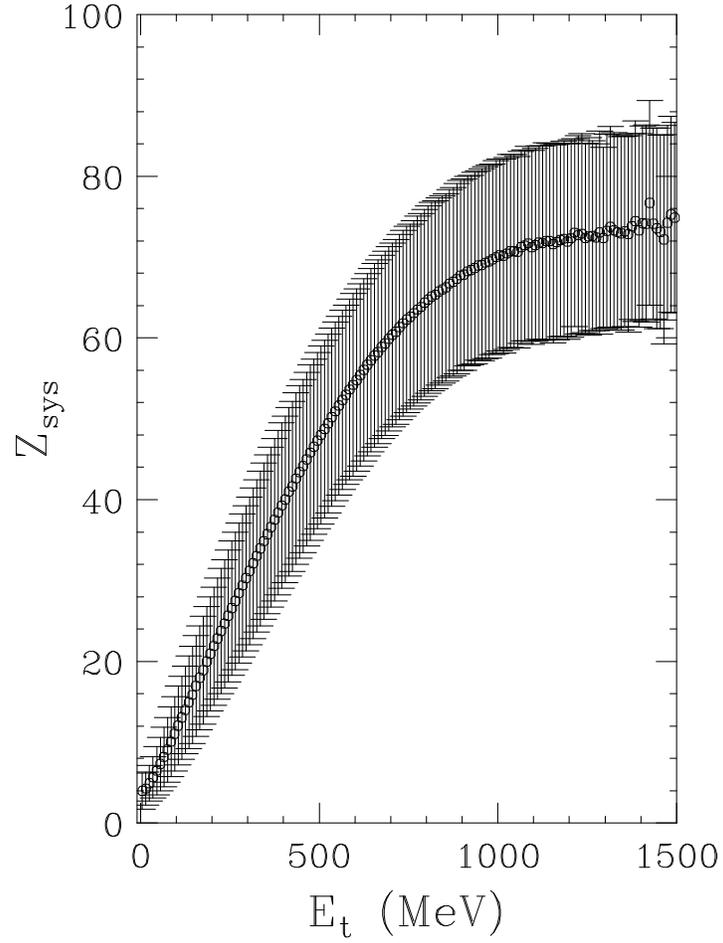}

\vspace*{3cm}

\caption[]{Dependence of the total charges of detected in a fragmentation
         event, $Z_{\rm sys}$, on the total transverse energy, $E_t$,
         detected in the experiment 55 $A$MeV Kr+Au \protect{\cite{Tsang}}. The
         error bars indicate the width (standard deviation) of the distribution
         on an event-by-event basis.}
\end{figure}

\clearpage

\begin{figure}
\vspace*{15cm}

\includegraphics{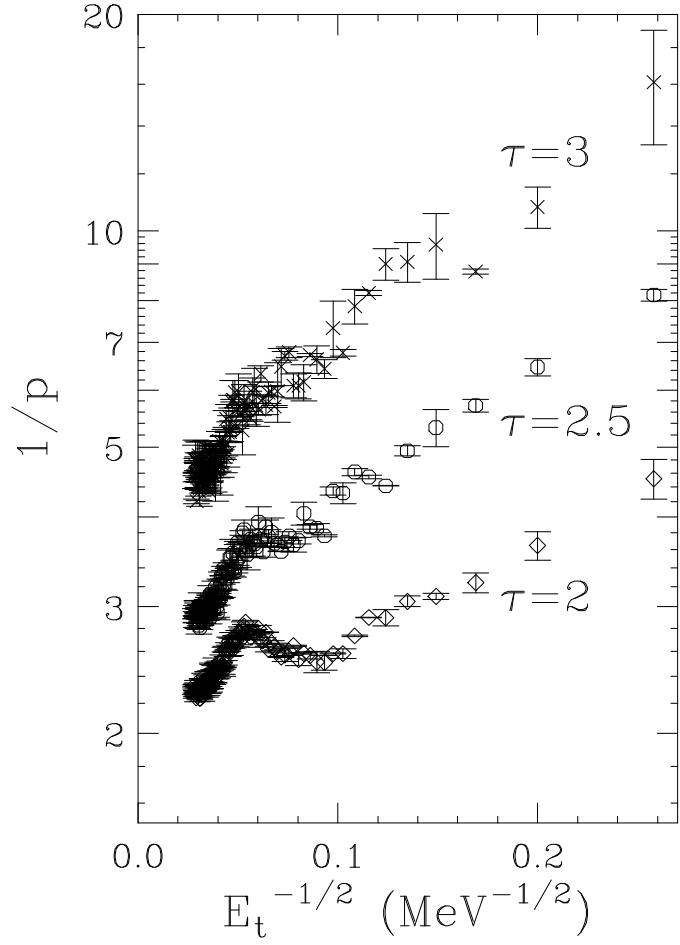}

\vspace*{3cm}

\caption[]{Dependence of the binomial parameter $p$ of the IMF distribution
on the transverse energy for three different values of $\tau$.}
\end{figure}

\clearpage

\begin{figure}
\vspace*{15cm}

\includegraphics{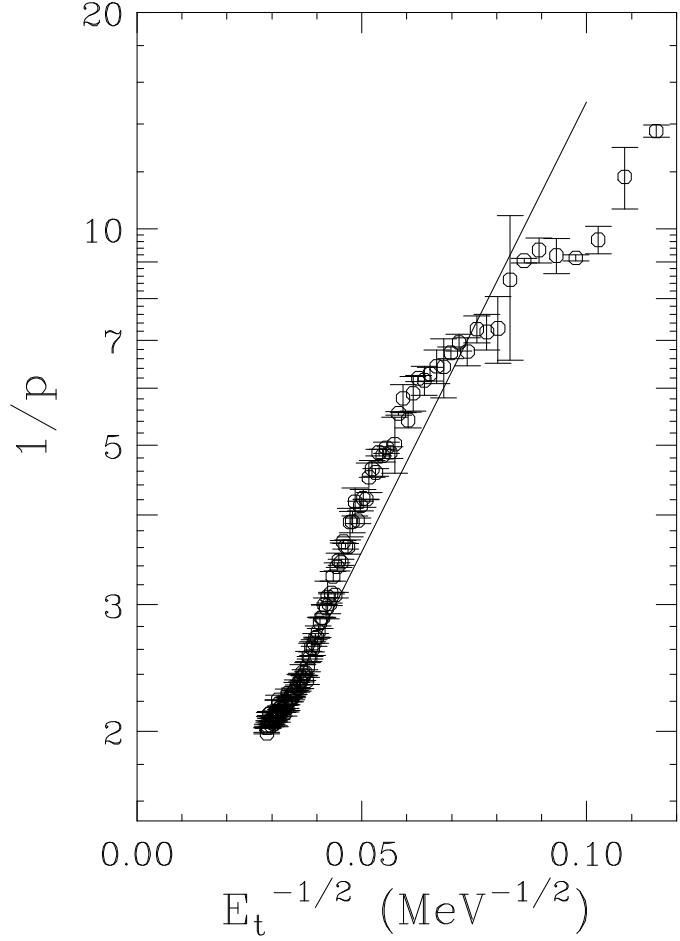}

\vspace*{3cm}

\caption[]{Dependence of the binomial parameter $p$ of the IMF distribution
on the transverse energy, assuming that the effective power $\tau$ of the 
fragment probability increases linearly with transverse energy. Plot symbols
with error bars represent our calculations, the solid line is a fit to
the experimental data of Moretto and collaborators.}
\end{figure}
\end{document}